\documentclass[aps,reprint,superscriptaddress]{revtex4-1}
\usepackage{graphicx}
\usepackage{hyperref}
\begin{document}

%\preprint{}

\title{Slow and fast light enhanced light drag in a moving microcavity}


\author{Tian Qin}
\author{Jianfan Yang}
\author{Fangxing Zhang}
\author{Yao Chen}
\affiliation{The State Key Laboratory of Advanced Optical Communication Systems and Networks, 
University of Michigan-Shanghai Jiao Tong University Joint Institute,
Shanghai Jiao Tong University, Shanghai 200240, China
}
\author{Dongyi Shen }
\author{\\Wei Liu}
\author{Lei Chen}
\affiliation{MOE Key Laboratory for Laser Plasmas and Collaborative Innovation Center of IFSA, 
Department of Physics and Astronomy, Shanghai Jiao Tong University, Shanghai 200240, China
}
\author{Yuanlin Zheng}
\affiliation{MOE Key Laboratory for Laser Plasmas and Collaborative Innovation Center of IFSA, 
Department of Physics and Astronomy, Shanghai Jiao Tong University, Shanghai 200240, China
}
\affiliation{Department of Electrical and Systems Engineering, Washington University in St. Louis, St. Louis, Missouri 63130, USA}
\author{Xianfeng Chen}
\affiliation{MOE Key Laboratory for Laser Plasmas and Collaborative Innovation Center of IFSA, 
Department of Physics and Astronomy, Shanghai Jiao Tong University, Shanghai 200240, China
}
\author{Wenjie Wan}
\email[Corresponding author: ]{wenjie.wan@sjtu.edu.cn}
\affiliation{The State Key Laboratory of Advanced Optical Communication Systems and Networks, 
University of Michigan-Shanghai Jiao Tong University Joint Institute,
Shanghai Jiao Tong University, Shanghai 200240, China
}
\affiliation{MOE Key Laboratory for Laser Plasmas and Collaborative Innovation Center of IFSA, 
Department of Physics and Astronomy, Shanghai Jiao Tong University, Shanghai 200240, China
}
%\thanks{}


\date{\today}

\begin{abstract}
Fizeau experiment, inspiring Einstein’s special theory of relativity, reveals a small dragging effect of light inside a moving medium. Dispersion can enhance such light drag according to Lorentz’s predication. Here we experimentally demonstrate slow and fast light enhanced light drag in a moving optical microcavity through stimulated Brillouin scattering induced transparency and absorption. The strong dispersion provides an enhancement factor up to $\sim 10^4$ , greatly reducing the system size down to the micrometer range. These results may offer a unique platform for a compact, integrated solution to motion sensing and ultrafast signal processing applications.
\end{abstract}

%\keywords{}

\maketitle

% References should be done using the  \cite, \ref, and \label commands
\section{Introduction}
It is well-known that the speed of light in a vacuum is the same regardless of the choice of reference frame, according Einstein’s theory of special relativity \cite{einstein1905electrodynamics}. However, once light in a moving medium, it can be dragged by the host medium, known as Fresnel drag effect \cite{fresnel1818influence}, which was first theoretically predicted by Fresnel in 1818 and experimentally verified by Fizeau experiment in a water-running tube interferometer in 1859 \cite{fizeau1851ahl}. Later, Lorentz pointed out the influence of dispersion on the light-dragging effect \cite{lorentz6electromagnetic}, which was also experimentally verified by Zeeman \cite{zeeman1914fresnel,zeeman1915fresnel}. All of these foundational works play a very important role in the development of Einstein’s special relativity \cite{norton2004einstein}. However, under the circumstance of low or none dispersion, such light-dragging effect is so insignificant that requires either long travelling path or large velocity of host media for any observable results, for example, a $1.5$-meter-long water tube with $2$ m/s velocity in the Fizeau experiment \cite{fizeau1851ahl}, and a $1.2$-meter-long glass rod moving at $10$ m/s in the Zeeman’s experiments \cite{zeeman1914fresnel,*zeeman1915fresnel}. Therefore, large light dragging requires strong dispersion to enhance the effect for compact systems and any practical applications such as inertia or motional sensing \cite{shahriar2007ultrahigh}. 

Previously, strong dispersion enhanced light drag has been observed in highly dispersive media such as atomic vapors \cite{strekalov2004observation,safari2016light}, cold atoms \cite{kuan2016large} and rare-earth doped crystals \cite{franke2011rotary}, where ultra-narrow linewidth atomic resonances can enormously enhance the drag effect up to $\sim 10^5$ times as compared to low dispersive media like water \cite{fizeau1851ahl} or glass \cite{zeeman1914fresnel,zeeman1915fresnel}, greatly reducing the system sizes down to the millimeter scale. For example, one theoretical proposal \cite{Carusotto2003} utilizing an electromagnetically induced transparent (EIT)  medium of strong normal dispersion to enhance the dragging effect has been realized both in hot atomic vapors \cite{strekalov2004observation,safari2016light} as well as cold atoms \cite{kuan2016large}. However, the rigid experimental conditions requiring vacuum and precise temperature maintaining in these atomic systems highly limit their practical applications. Meanwhile, in solid-state systems, photonic crystal structures \cite{soljavcic2002photonic,vlasov2005active}, optomechanical devices \cite{weis2010optomechanically,safavi2011electromagnetically,bahl2011stimulated,Dong2015,Kim2015} and doped crystals \cite{turukhin2001observation,bigelow2003superluminal} have shown a great potential to manage optical dispersion in a compact form, allowing exotic propagation such as slow light and fast light \cite{boyd2001slow}. And thanks to the ultrahigh quality factors of resonances in these systems, the dispersion enhancement is comparable to those in atomic platforms, providing a fertile ground to test light-dragging effect. Particularly, both the normal dispersive EIT and the anomalous one associated with electromagnetically induced absorption (EIA) have been demonstrated in a micro-scale optomechanical microcavity \cite{Kim2015}. Although such dispersion enhanced light dragging effects have only been realized  in the normal dispersion regime so far, how the strong anomalous dispersion affects the light drag is still a critical question especially for the key gyroscopic application, which is still under the debate whether such dispersion enhancement can provide extra sensitivity \cite{shahriar2007ultrahigh,terrel2009performance}.

In this work, we theoretically and experimentally demonstrate strong dispersion enhanced light dragging effect arising from both slow and fast light induced by stimulated Brillouin scattering (SBS) processes in the same solid-state microcavity platform. In a high-quality (Q) microcavity, both acoustical and optical waves can be trapped resonantly in whispering gallery modes, the high-Q acoustical resonances induce sharp perturbations in those optical ones, creating electromagnetically induced transparency or absorption depending on their relative phase. In these spectral disturbed regimes, optical dispersion relations are strongly altered resulting in slow or fast group propagation of light. Based on that,  Lorentz’s predications of light drag are examined with an enhancement factor up to $10^4$ in both dispersive regimes through a fiber coupled microcavity setup in a vibrating mode. Moreover, such optomechanical micro-devices allow us to actively tune the induced dispersion. As a result, the light dragging can also be modified accordingly, further validating the nature of dispersion-related enhancement factor. At last, a Fano-like resonance obtained by detuning technique in the microcavity permits the observation of the drag by both slow and fast light in close spectral proximity. These results extend the special relativity experiment to a new solid-state system at microscale, opening up new avenues for their practical applications in motion sensing, ultrafast signal processing and even for on-chip time reciprocity experiments.

\section{Theory of Light Drag in a Dispersive Moving Microsphere Cavity}
\begin{figure*}
\centering
\includegraphics[width=0.935\linewidth]{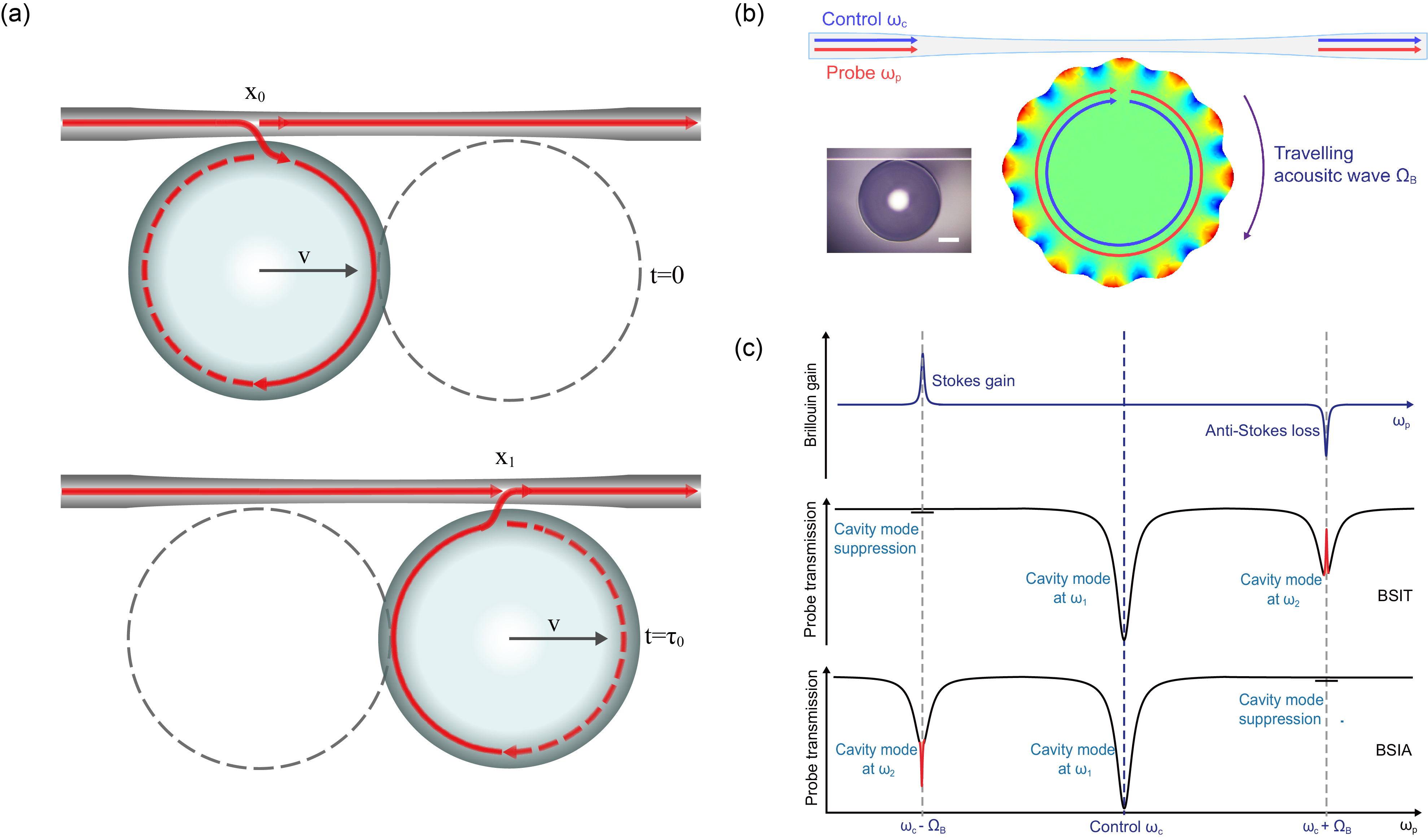} 
\caption{Light-dragging effect in a EIT/EIA microsphere cavity. (a) Schematic illustration of the light-dragging effect in a moving microsphere cavity. The microsphere cavity moves parallel to the tapered fiber at velocity $v$ with respect to the stationary laboratory frame. In this frame, light is coupled into the cavity at position $x_0$ and then coupled out of the cavity at position $x_1$ after finishing a round trip in the cavity. The intra-cavity light is dragged by the moving medium, resulting a light-dragging shift of round-trip phase.  (b) Schematic illustration of forward Brillouin scattering excited by a strong control laser. The control and the probe laser are coupled to two phase-matched cavity modes  and interact with each other via a travelling acoustic wave. The microscope image of the coupling system is showed in the inset, with a scale bar of $50\mu$m. (c) Spectrum diagram for EIT and EIA induced by forward Brillouin scattering. For EIT, probe laser hits the higher-frequency cavity mode while Stokes gain is suppressed. The probe experiences anti-Stokes loss inside the cavity, which leads to the slow light effect observed at the through port. For EIA, probe laser hits the lower-frequency cavity mode while anti-Stokes gain is suppressed.  The probe experiences Stokes gain inside the cavity, which leads to the fast light effect observed at the through port.\label{fig:schematic}}
\end{figure*} 

Our work to investigate the light dragging effect is based on a tapered fiber coupled microsphere resonator, which vibrates translationally parallel to the coupling fiber to provide the drag [Fig.~\ref{fig:schematic}(a)]. In this configuration, light can tunnel from the input fiber into the microcavity to form a whispering gallery mode (WGM) due to total internal reflection \cite{knight1997phase}. Such tunneling occurs in both directions, allowing light inside the microcavity to be coupled back to the fiber as well. During the light’s resonant staying inside the microcavity, the cavity has already travelled translationally at some distance, effectively inducing the light drag. However, such dragging effect is relatively small, for example, Carmon's team has shown that a millimeter size resonator with $Q_{\text{optical}}\sim 10^6$ spinning at $\sim 6000$ round per second can only cause $\sim 20$ Mhz resonance shift \cite{Maayani2018}. In order to enhance the process, strong dispersion must be considered. It is well-known that extraordinary dispersion can be found in EIT or EIA \cite{Harris1992,lezama1999electromagnetically}, where two spectral adjacent resonances are coherently coupled to each other, giving rise to the strong disturbance in the phase spectrum and causing the strong dispersion relationship nearby. Such phenomenon is not only limited to atomic systems, but also in optical microcavities, e.g. even two coupled microcavities allow EIT-like transmission of light \cite{xu2006experimental}. Moreover, when increasing optical intensity in  microcavities, the enhanced nonlinearity can stimulate other resonances such as mechanical \cite{weis2010optomechanically}, acoustical \cite{bahl2011stimulated,Dong2015,Kim2015} and optical ones \cite{zheng2016optically}, these resonances can also interplay and couple with the pump resonance to form an EIT or EIA spectrum. As a result, one associated consequence of such strong dispersion permits the observation of slow light or fast light in terms of group velocity \cite{Kim2015}. This opens up a unique window for us to test Fresnel light drag. As shown in Fig.~\ref{fig:schematic}(b), our microsphere cavity can support both optical and acoustic whispering-gallery-type resonances, where a strong optical pump light is launched into one optical WGM, simultaneously generating a Stokes photon with an extra acoustic phonon, or an anti-Stokes photon via absorbing one phonon through an electrostriction-induced stimulated Brillouin scattering (SBS) process \cite{bahl2011stimulated,Kim2015,Dong2015}. The Stokes or anti-Stokes processes occur depending on whether an extra optical resonance can be found  for the scattering photon as shown in Fig.~\ref{fig:schematic}(c). This leads to EIA or EIT  transmission respectively due to inherent loss or gain mechanisms \cite{fleischhauer2005electromagnetically}. Meanwhile, it has been reported slow and fast light can reach up to microsecond level \cite{Kim2015},  this can hugely enhance the light drag according to Lorentz’s predication \cite{lorentz6electromagnetic}. 

The mechanism of slow and fast light enhanced light drag in a moving WGM microcavity can be considered by applying Lorentz transform to the coupling system. Here we assume that light is coupled into a microcavity moving along the tapered fiber at velocity $v$ [see Fig.~\ref{fig:schematic}(a)], the round-trip time and the frequency of light with respect to the frame of moving cavity $(S')$ are given by the Lorentz transform:
\begin{eqnarray}
\tau'_0&&= \frac{Ln(\omega')}{c}, \label{eq:tau}\\
\omega'&&= \omega \left( 1-\frac{v}{c}\right)\label{eq:omega},
\end{eqnarray}
where $L$ is the perimeter of the cavity, $n(\omega')$ is the refractive index of the cavity in the moving frame. Therefore, the round trip phase of light inside the cavity is modified by the drag of the medium, the resulting phase shift compared to the stationary situation is obtained from Eq.~(\ref{eq:tau}) and Eq.~(\ref{eq:omega})
\begin{equation}
\Delta\phi_{\text{drag}}(\omega)\simeq \frac{L\omega}{c}\left[ n(\omega')\left(1-\frac{v}{c}\right)-n(\omega)\right]. \label{eq:dphi_single}
\end{equation}
As $v$ is much smaller than $c$ in general case, the light-dragging phase shift $\Delta\phi_{\text{drag}}$ (internal) is expected to be very small unless large dispersion is present. To achieve large dispersion by means of slow or fast light, a weak probe laser at $\omega_p$ is scanned through the Stokes or anti-Stokes resonances induced by a strong control laser at $\omega_c$ with SBS as shown in Fig.~\ref{fig:schematic}(b)(c), which is similar to prior works \cite{Kim2015,Dong2015}. Then the probe laser can experience large dispersion induced by EIT/EIA near $\omega_c\pm \Omega_B$. The refractive index of the probe under the moving frame $n(\omega_p')$ can be described by
\begin{eqnarray}
n(\omega_p')=n(\omega_p)+\frac{v}{c}(\omega_c-\omega_p)\cdot\frac{n_g(\omega_p)-n(\omega_p)}{\omega_p}. \label{eq:refractive_index_2}
\end{eqnarray}
Note that the strong group dispersion term $n_g(\omega_p)$  tremendously enhances the relativity effect brought by the velocity $v$, similar to the light drag in the atomic vapors and cold atoms \cite{safari2016light,kuan2016large}. But the control laser is also dragged by the medium, which is reflected in the cancellation term, i.e. $\omega_c-\omega_p$, discussed in Sec.~IA of the Supplemental Material \cite{supplemental}. By substituting $n(\omega')$ in Eq.~(\ref{eq:dphi_single}) with Eq.~(\ref{eq:refractive_index_2}) and considering the dragging effect of the strong laser \cite{Carmon2004,supplemental}, the relative light-dragging phase shift of the probe can be obtained as
\begin{eqnarray}
\Delta\phi_{\text{drag\_p}}(\omega_p)\simeq && \frac{L(\omega_c-\omega_p) n(\omega_c)v}{c^2}\left[ 1+\frac{n_g(\omega_p)-n(\omega_p)}{n(\omega_c)}\right] \nonumber\\
\equiv &&\frac{L(\omega_c-\omega_p)n(\omega_c)v}{c^2}F_d(\omega_p), \label{eq:dphi2_2}
\end{eqnarray}
where $F_d (\omega_p)$ is defined as the light-dragging enhancement factor, which is approximately proportional to $n_g(\omega_p)$ when $n_g(\omega_p)\gg n(\omega_p)$. Since $n_g(\omega_p)$ inside the cavity is found to be negatively proportional to the probe time delay at the through port of the tapered fiber \cite{Kim2015,Huet2016} (negative sign due to the cavity out-coupling, details in Sec.~ID of the Supplemental Material \cite{supplemental}), tremendous enhancement of light drag is expected with the presence of fast/slow light.
An enhancement up to $\sim -10^4$ can be expected based on the observation of $n_g \sim -1.6\times 10^4$ in the case prior works of EIT/EIA \cite{Kim2015}, slightly smaller than the cold atom case \cite{kuan2016large} in term of magnitude. In our experiments, the transmission of the probe through the fiber coupled microcavity system is altered by such light-dragging induced phase shift through interfering with the residual light inside the fiber \cite{gorodetsky1999optical}, giving us a tool to measure the dragging phase shift (see Sec.~IB of the Supplemental Material \cite{supplemental})
\begin{eqnarray}
\Delta T_{\text{drag\_p}}(\omega_p)\simeq &&\frac{2a\kappa_2(1-a^2)(1-\kappa_2^2)\Delta\phi_0(\omega_c\pm \Omega_B)}{[(1-a\kappa_2)^2+a\kappa_2\Delta\phi_0^2(\omega_c\pm \Omega_B)]^2}\nonumber\\
&&\times \Delta \phi_{\text{drag\_p}}(\omega_p) \label{eq:deltaT2_2}\\
\equiv&& C \cdot \Delta\phi_{\text{drag\_p}}(\omega_p) \label{eq:deltaT2},
\end{eqnarray}
where $a$ is the round-trip absorption loss of the cavity, $\kappa_2$ is the coefficient of transmission between the tapered fiber and the cavity, $\Delta\phi_0(\omega_c\pm\Omega_B)$ is the initial round-trip phase mismatch of the cavity at anti-Stokes/Stokes frequency, depending on the case of slow light($+$) or fast light($-$). It is convenient to define a parameter $C$ to characterize the effect of small phase perturbation on the transmission near an optical resonance. Then the phase shift induce by dispersion enhanced light-dragging effect can be converted into the change of probe transmission. 
\section{Experimental Results and Discussion}
\subsection{Light-dragging Effect in an EIT Microcavity}

\begin{figure}
\centering
\includegraphics[width=0.95\linewidth]{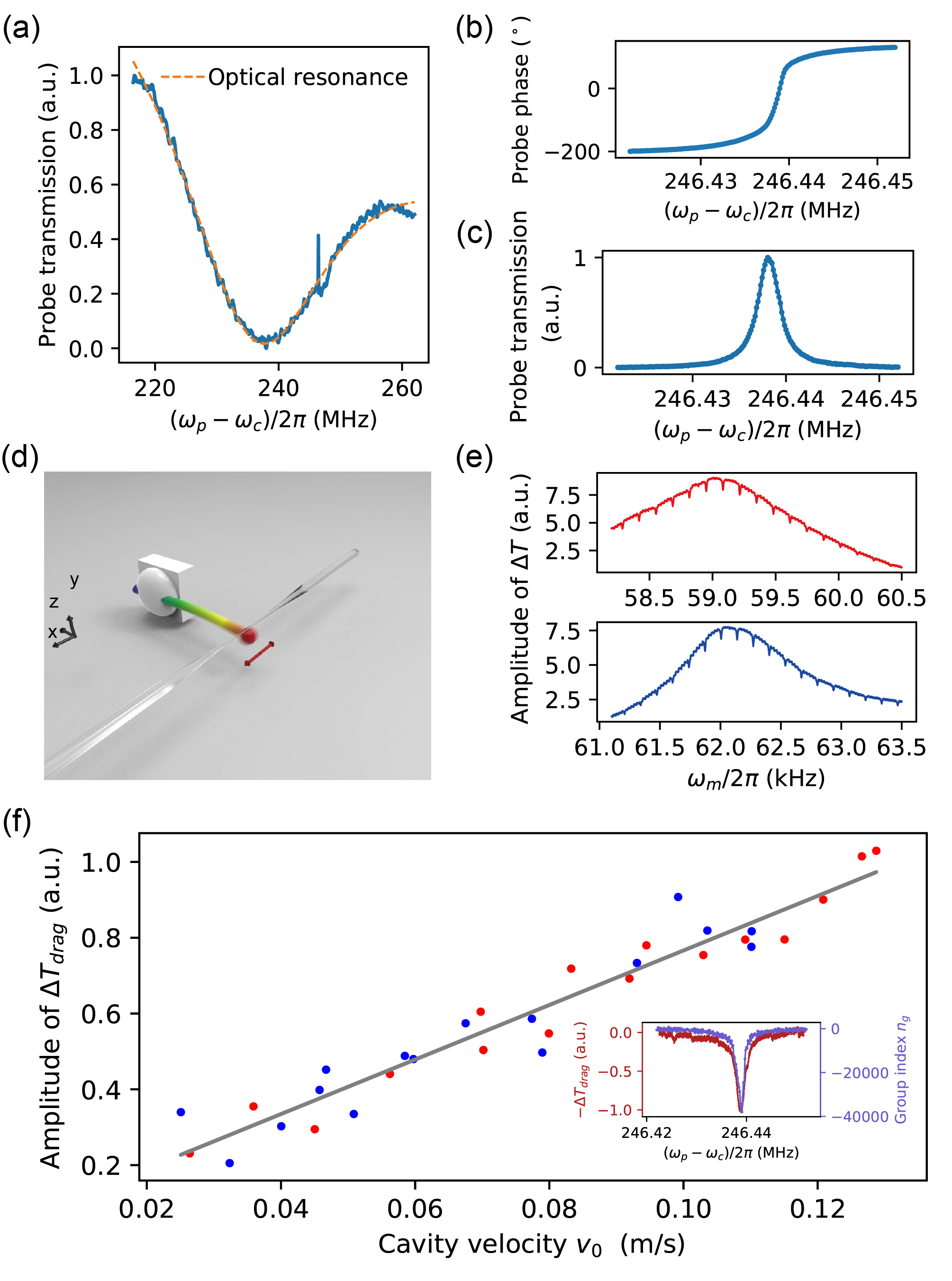} 
\caption{Light-dragging effect in a EIT microcavity. (a) Transparency is observed in a higher-frequency cavity mode ($Q=6.4\times 10^6$). (b) Large positive slope in probe phase takes place at $\omega_p=\omega_c+\Omega_B$, indicating strong normal dispersion. (c) Probe time delay of $338$ $\mu\text{s}$ is observed in the tapered fiber, corresponding to a intracavity group index of $n_g=-4.7\times 10^4$ (see Sec.~ID of the Supplemental Material \cite{supplemental}) and an light-dragging enhancement factor $F_d=-3.2\times 10^4$. (d) Schematic illustration of a vibrating microsphere cantilever driven by a piezoelectric vibrator. (e) The amplitude of perturbation on probe transmission led by the vibrating cavity at $59$ kHz and $62$ kHz cantilever mode resonances, which is measured by a lock-in amplifier with probe frequency scanning near the EIT window ($20$kHz range and $4$Hz repetition rate) and piezo-driving frequency scanning near the mechanical resonances ($2$kHz range and $10$mHz repetition rate) simultaneously. The perturbation is contributed by both light-dragging effect and coupling-gap modulation. Small light-dragging induced dips are observed on humps caused by coupling-gap modulation. (f) Amplitude of the light-dragging induced perturbation on probe transmission, $\Delta T_{\text{drag}}$, as a function of the amplitude of vibration velocity $v_0$. The inset shows a single dip and corresponding dispersion profile, indicating the dependence of $\Delta T_{\text{drag}}$ on group index $n_g$. \label{fig:BSIT}}
\end{figure} 

Firstly, we realize EIT by Brillouin scattering induced transparency through an anti-stokes process in our optomechanical microcavity (Detailed experimental setup showed in Sec.~IIA of the Supplemental Material \cite{supplemental}). Fig.~\ref{fig:BSIT}(a) depicts an ultra-narrow mechanical resonance peak ($3$ kHz linewidth) inside an optical cavity mode ($30.4$ Mhz linewidth), similar to the previous works  \cite{Kim2015,Dong2015}. Its phase spectrum in Fig.~\ref{fig:BSIT}(b) exhibits a rapid climb near the center of the induced transparency window, which later will be explored for the enhancement for studying the light drag. In the meantime, we observe a slow light of group time delay up to $338$ $\mu\text{s}$ [see Fig.~\ref{fig:BSIT}(c)] by probing a secondary light through EIT window, which can significantly amplify the light drag according to the theory above. Based on this platform, we exam dispersion enhanced light drag by moving the microcavity along the coupling fiber while maintaining a constant coupling gap in a vibrating manner such that the relative speed is sinuously modulated (see Sec.~IC of the Supplemental Material  \cite{supplemental}). 
Here the microsphere is fabricated on the top of an optical fiber, effectively composing a cantilever structure which oscillates at several natural mechanical resonances \cite{li2016simultaneous} as shown in the illustration [see Fig.~\ref{fig:BSIT}(d)]. We choose two of these resonances (at $59$ kHz and $62$ kHz) driven by an attached piezoelectric vibrator [see Fig.~\ref{fig:BSIT}(e)] for maximizing moving speed to exam any observable dragging effect. This is done by reading out the transmission of the probe laser whose frequency is periodically ($4$ Hz) scanned through one slow-light window (in $20$ kHz frequency range), in the meantime, the microcavity cantilever’s vibration frequency is slowly swept through those two mechanical resonances (with repetition rate of $10$ mHz in frequency range of $2.5$ kHz). As a result, small dips in the envelope of probe transmission are observed periodically on a slow-varying background humps [see Fig.~\ref{fig:BSIT}(e)], which can be explained by the superposition of light-dragging effect (dips) and coupling-gap modulation (humps) (see Sec.~IC of the Supplemental Material\cite{supplemental}).  These dips are measured through a lock-in technique recording the temporal sinusoidal modulation induced by the light drag of the vibrating cantilever. In the ideal case, the coupling gap between the microcavity and the fiber is kept constant when they are perfectly aligned. However, slight misalignment can lead to a huge response in the probe transmission exhibiting a background hump. To confirm the origin of the humps and calibrate the  velocity of the cavity, the aligned and misaligned motion of the vibrating microcavity are separately measured to give  a maximum vibration velocity of $0.12$ m/s accompanied with $10$ nm change in coupling gap by monitoring the change of the WGM coupling efficiency induced by the coupling gap with perpendicular alignments of the piezoelectric vibrator (see Sec.~IIB of the Supplemental Material \cite{supplemental}). 

Such dips in the envelope of transmission are caused by the interference between the direct transmitted light and those dragged by the microcavity during the coupling processes (see Sect.~IIC of the Supplemental Material \cite{supplemental}). Their height directly measures the dragging phase due to destructive or constructive interferences according to Eq.~(\ref{eq:deltaT2}). Under the condition of slow-light enhancement, the amplitude of light-dragging perturbation,  measured from the height of the dips linearly scale with the velocity of the microcavity [see Fig.~\ref{fig:BSIT}(f) for the two cases in Fig.~\ref{fig:BSIT}(e)], which is calibrated by  the displacement and the frequency of the vibration. This confirms the velocity dependence of light drag in Eq.~(\ref{eq:dphi2_2}). Meanwhile, since the dragging effect is linearly enhanced by the strong dispersion in slow light, the profile of a dip is compared with the calculated group index near the slow light regime in the inset of Fig.~\ref{fig:BSIT}(f). The light-dragging enhancement factor maximizes its amplitude in the center, where the transmission dip led by light drag also reach its minimum. Meanwhile, the linewidths of these two terms also match well with each other, indicating a tunable enhancement by small detuning is possible in the regime of interest.

\subsection{Light-dragging Effect in an EIA Microcavity}
\begin{figure}
\centering
\includegraphics[width=\linewidth]{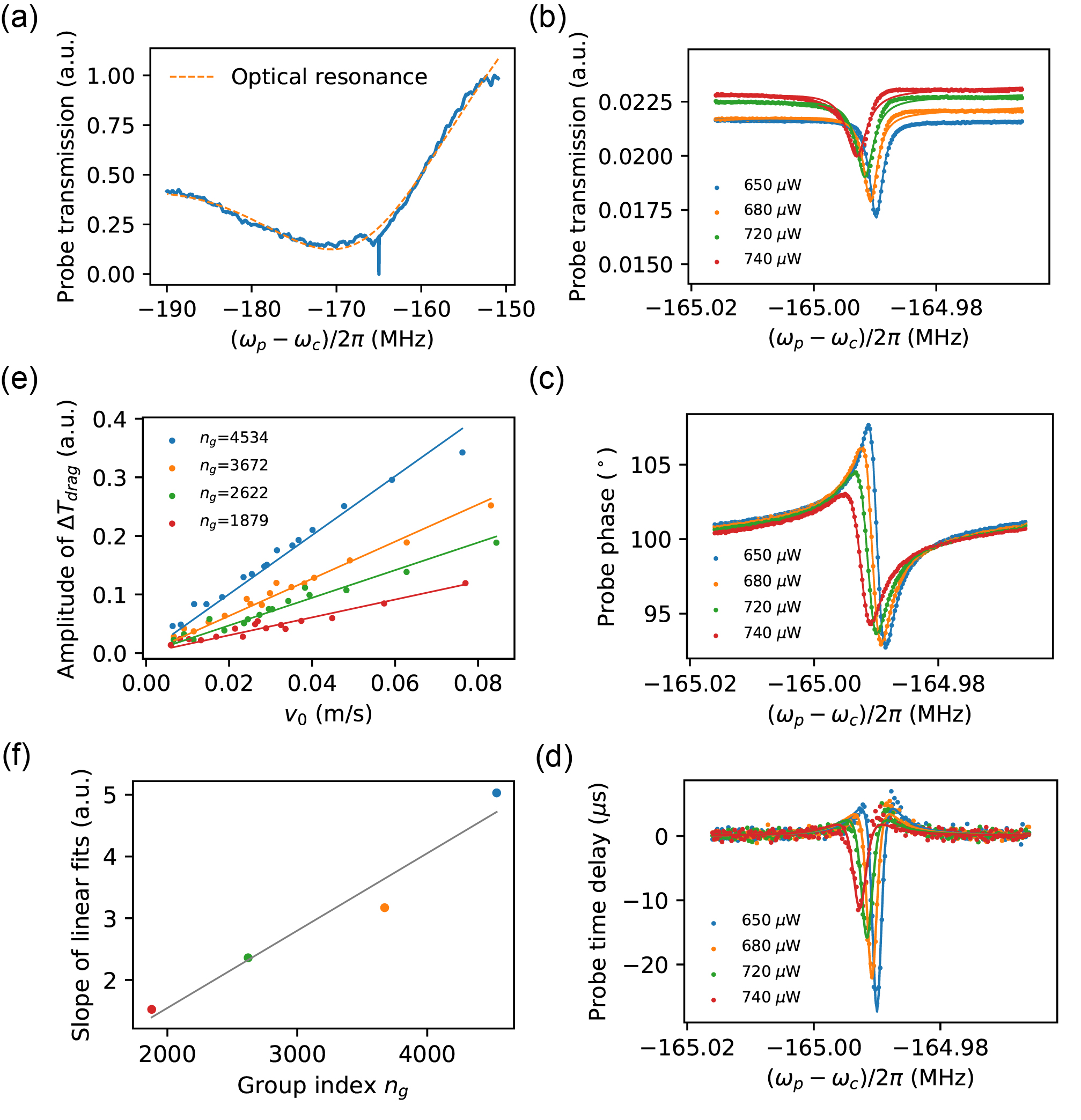} 
\caption{Light-dragging effect in a EIA microcavity. (a) Absorption is observed in a lower-frequency cavity mode ($Q=5.4\times 10^6$) with $650$ $\mu\text{W}$ control power. (b) The probe transmission near the absorption window at $\omega_c-\Omega_B$, the linewidth of the transmission profile decreases with increasing control power. (c) The probe phase response shows large negative slope at the center of the absorption window and weaker negative slope on two wings. The slope deceases with increasing control power. (d) The probe time advancement reaches $26$ $\mu\text{s}$ at the center of the absorption window with $650$ $\mu\text{s}$ control power, corresponding to an intracavity group index of $4534$ (see Sec.~ID of the Supplemental Material \cite{supplemental}) and enhancement factor of $3.1\times 10^3$. The probe time advancement decreases with increasing control power, which implies decreasing intracavity group index of $4534$(blue), $3672$(orange), $2622$(green) and $1879$(red). (e) Amplitude of the light-drag induced perturbation on probe transmission, $\Delta T_{\text{drag}}$, is plotted as a function of velocity $v_0$. Experiment results (dots) taken with different group index $n_g$ are labeled with different colors. Light-dragging enhancement factor can be compared via the slope of linear fits (solid lines). (f) Normalized enhancement factor, represented by the slopes in (b), is plotted as a function of group index $n_g$.  The solid line represents a linear fit. \label{fig:BSIA}}
\end{figure} 

Anomalous dispersion can induce strong enhancement to light drag as well. Fig.~\ref{fig:BSIA} shows that light drag enhancement can also result from the anomalous dispersion in a Stokes SBS configuration inside the microcavity. The complex phase of the Stokes induced acoustical resonance coherently interferes with one optical resonance, resulting in EIA at the through port in Fig.~\ref{fig:BSIA}(a)(b) \cite{Kim2015}. Unlike the previous case of EIT, the phase spectrum of EIA has large negative slope at the center of the absorption dip, exhibiting a near-Lorentzian lineshape [see Fig.~\ref{fig:BSIA}(c)]. Note that such phase relation composes both contribution from the acoustical resonance and the optical one. Two positive dispersive wings in the phase spectrum can also enhance the light drag, which shall be elaborated later. The central strong dispersion leads to fast light phenomena measured by the group phase advance instead of group delay up to $26$ $\mu\text{s}$ as shown in Fig.~\ref{fig:BSIA}(d).  The resulting dragging effect also induces dips in the envelope of perturbation on probe transmission. Note that, similar to the previous case of EIT, such dips mix up the coherent interference from both optical resonances, the acoustical resonance as well as light drag (see Sec.~IIC of the Supplemental Material \cite{supplemental}). The height of the dips can be translated to the light-dragging perturbation, which is characterized to be linearly dependent on the corresponding microcavity velocity [see Fig.~\ref{fig:BSIA}(e)], manifesting the nature of speed-dependent light drag. Previously, only EIT was considered for the light dragging medium for its feature of low loss \cite{kuan2016large}. Here we have demonstrated enhanced light drag in a EIA medium, which is enabled by relatively small  absorption but large phase disturbance from the dispersion, i.e. the key factor for light-dragging enhancement.

Moreover, we illustrate that such dispersion enhanced light drag can be tunable through a dispersion management as shown in Fig.~\ref{fig:BSIA}. In our experiments, the linewidth of EIA windows can be effectively controlled by the pumping power as reported by Dong's team \cite{Dong2015}. For the same EIA mode, when gradually increasing the pumping power from $650$ $\mu\text{W}$ to $740$ $\mu\text{W}$, the depth of the transmission dips decreases by around $50\%$ due to nonlinear phase matching condition in SBS \cite{Dong2015}. More importantly, their linewidths increase [see Fig.~\ref{fig:BSIA}(b)], resulting in smoother phase jumps [see Fig.~\ref{fig:BSIA}(c)] and reducing the group time advances by half [see Fig.~\ref{fig:BSIA}(d)]. As a result, the corresponding light drag measured under these pumping conditions now exhibit slightly different slopes with respect to the cavity velocity [see Fig.~\ref{fig:BSIA}(e)], since the dragging enhancement factor $F_d$ has been modified by different dispersion features in these configurations. For verification purpose, we also plot these slope coefficients from Fig.~\ref{fig:BSIA}(e) as a function of corresponding group indexes in Fig.~\ref{fig:BSIA}(f), which shows a good linearity confirming the linear dependence of light-dragging enhancement factor on the dispersion relation, i.e. group index as shown in Eq.~(\ref{eq:dphi2_2}).

\subsection{Light-dragging Effect with Normal and Anomalous Dispersion}
\begin{figure}
\centering
\includegraphics[width=.95\linewidth]{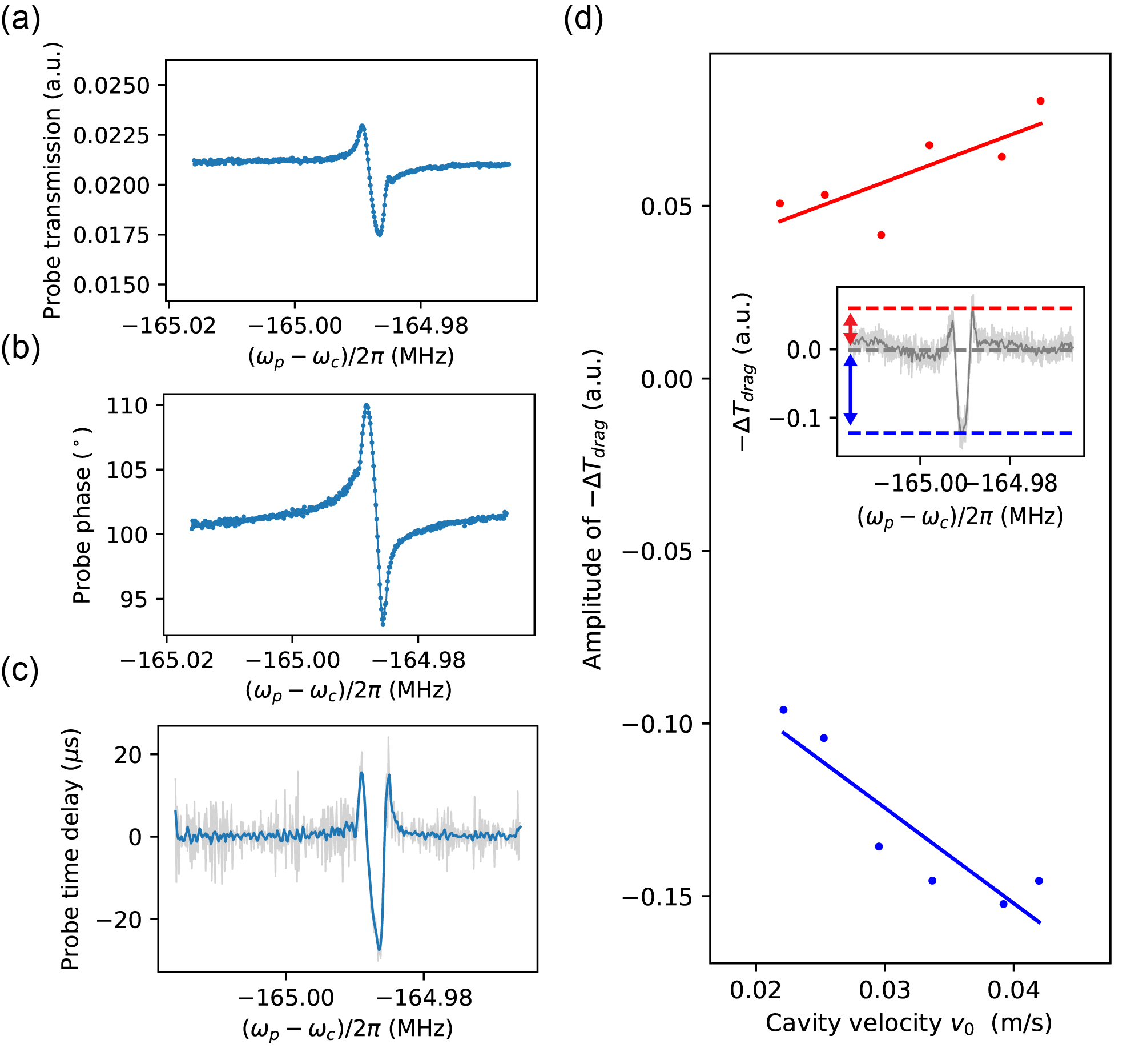} 
\caption{Light-dragging effect in a microcavity with Fano-like EIA window. (a) A Fano-like transmission diagram is observed near $\omega_p=\omega_c-\Omega_B$ with detuned control frequency. (b)The probe phase response shows negative slope at the center, while the positive slope wings are much steeper comparing to the previous case of EIA. (c) The negative and positive slopes in the probe phase response are corresponding to probe time delay and advancement. The maximum delay and advancement can be translated to intracavity group index of $-2591$ and $4563$ respectively. (d) Amplitude of the light-dragging induced perturbation on probe transmission, $\Delta T_{\text{drag}}$, is plotted as a function of cavity vibration velocity with both slow-light and fast-light enhancement, where the enhancement of negative (red) and positive (blue) intracavity group index are showed by the slope of linear fits. \label{fig:distort}}
\end{figure} 

At last, normal and anomalous dispersion enhanced light-dragging effect are demonstrated simultaneously in a SBS induced Fano-like transmission spectrum in Fig.~\ref{fig:distort}. Inspired by the recent experiments including ours \cite{Dong2015,zheng2016optically} to obtain the Fano resonance by slightly detuning the pump off the center of optical resonance in EIT, we realize such Fano-like transmission using the same technique, where the asymmetric Fano lineshape arises from the interference between two coupled resonances (optical one and acoustic one) and a coherent background \cite{zheng2016optically,Boiler1991,lezama1999electromagnetically}. Such Fano resonance has a distinct asymmetric lineshape in the intensity spectrum [see Fig.~\ref{fig:distort}(a)] as well as the phase spectrum  [see Fig.~\ref{fig:distort}(b)], which contains highly normal dispersive regime on both wings but anomalous one in the center. As a result, the probe time delays are present on both wings, while shifting to probe time advancement in the center [see Fig.~\ref{fig:distort}(c)].  This opens up an opportunity for us to observe the light drag through these dispersion relations at the same time. The measured enhanced light drag now behaves oppositely to each other in term of slope, with a ratio around $-0.65$. In the meantime, the intracavity group index is estimated to be $-2591$ for the slow light regime and $4563$ for the fast light regime calculated from the probe time delay/advancement in Fig.~\ref{fig:distort}(c). The ratio of group index in the two situations, calculated to be $-0.57$,  coincides with the slope ratio above. This is an important result since the dispersion enhanced light drag can be easily flipped by a frequency detuning less than the linewidth  of the resonance (acoustical one), for example, one can purposely position two spectrally close probes in these two opposite dispersion regime to maximize their beating signals for the optical gyroscope \cite{kadiwar1989optical,li2017microresonator}; alternatively, two counter-propagating waves, if presented with proper central frequency, can experience the two opposite drag enhancement in a spinning setup like previous work \cite{Maayani2018}, greatly reducing the system size and spinning speed requirement for optical isolator. This also opens up new possibility for manipulating dispersion enhanced light drag for applications in sensing and spectroscopy.   

\subsection{Discussion}
In summary, we have theoretically and experimentally exam Lorentz’s prediction of dispersion enhanced Fresnel light drag in both slow and fast light configurations. Compared with prior works, our new platform is implemented on a solid-state microcavity with a micrometer dimension and a speed sensitivity down to cm/s, thanks to a $\sim 10^4$ dispersion enhancement factor introduced by the stimulated Brillouin scattering process inside the optical microcavity. Both slow and fast light enhanced light dragging effects have been demonstrated with a flexible tunability from active pumping or detuning, paving a way for related and highly demanded applications in inertial, gyroscopic motional sensing. The demonstrated compact system also offers a new avenue for all-optical and on-chip photonics applications. 
\begin{acknowledgments}
This work was supported by National key research and development program (Grant No. 2016YFA0302500, 2017YFA0303700); National Science Foundation of China (Grant No. 11674228, No. 11304201, No. 61475100); Shanghai MEC Scientific Innovation Program (Grant No. E00075); Shanghai Scientific Innovation Program (Grant No. 14JC1402900); Shanghai Scientific Innovation Program for International Collaboration (Grant No. 15220721400).
\end{acknowledgments}

%\bibliography{reference}
%
\end{document}